\documentclass[12pt]{article}
\usepackage{color}
\usepackage{stmaryrd}
\usepackage{bbm}
\usepackage{mathrsfs}
\usepackage{amsfonts}
\usepackage{amssymb}
\usepackage{amsmath}
\usepackage{amsthm}
\usepackage{cases}
\usepackage{indentfirst}
\usepackage{graphicx}
\usepackage{algorithm}
\usepackage{amsthm,amsmath,amssymb}
\usepackage{mathrsfs}
\usepackage{algorithmic}
\usepackage[sort&compress,numbers]{natbib}
\usepackage{amsmath,amssymb,mathrsfs,amscd,graphicx,float,subfigure}

\usepackage{chngpage}
\usepackage{array}

\newtheorem{remark}{Remark}
\allowdisplaybreaks[4]

\def\be{\begin{eqnarray}} \def\ee{\end{eqnarray}} 
  \def\({\left(} \def\){\right)}
\def\bc{\begin{center}} 
\def\ec{\end{center}}  
\def\bey{\begin{eqnarray*}}\def\eey{\end{eqnarray*}}

\begin{document}

\title{ {\bf Some Wadati-Konno-Ichikawa type  integrable systems and their constructions}
\footnotetext{* Corresponding author.  }}
\author{Shou-Feng Shen, \quad Guo-Fang Wang, \quad Yong-Yang Jin, \quad   Xiao-Rui Hu $^{*}$\\}
\date{}
\maketitle
\begin{center}
\begin{minipage}{135mm}
\noindent {\small  Department of Applied Mathematics, Zhejiang
University of Technology, Hangzhou 310023, China}\\
\end{minipage}
\end{center}

\noindent{\bf Abstract:}

 A standard-form Wadati-Konno-Ichikawa(WKI) type integrable  hierarchy  is derived from a corresponding matrix
spectral problem associated with the Lie algebra $sl(2,\mathbb{R})$. Each equation in the resulting
hierarchy has a bi-Hamiltonian structure furnished by the trace identity.
Then, the higher grading affine algebraic construction of some special cases is proposed.
 We also show that generalized short pulse equation arises naturally from the negative WKI flow.

 \vspace{0.5cm}

\noindent{\bf Keywords:} WKI integrable  hierarchy;  bi-Hamiltonian structure; higher grading structure;  short pulse equation.

 \vspace{0.5cm}

\noindent{\bf PACS numbers:} 02.30.Ik

\noindent{\bf MSC numbers:} 37K05; 37K10; 35Q53
 \vspace{0.5cm}

 \numberwithin{equation}{section}

\section{Introduction}

Integrable systems (or integrable hierarchies) have attracted extensive attention in natural science because of successful description and explanation of nonlinear phenomena. Matrix spectral problems associated with Lie algebras are crucial keys to construct
integrable hierarchies. There has been a lot of work on the generation of integrable hierarchies from matrix spectral problems
and interesting examples contain the Ablowitz-Kaup-Newell-Segur (AKNS) hierarchy, the Kaup-Newell (KN) hierarchy,
the Wadati-Konno-Ichikawa (WKI) hierarchy, the Korteweg-de Vries (KdV) hierarchy, the modified KdV hierarchy, the Benjamin-Ono hierarchy, the  Boiti-Pempinelli-Tu (BPT) hierarchy, the Dirac hierarchy and the coupled Harry-Dym hierarchy \cite{akns,kn, wki, ff, aks, bpt, af1, af2}.  These integrable hierarchies usually possess nice properties such as having hereditary recursion operators, being multi-Hamiltonian, and carrying infinitely many commuting symmetries and conservation laws. The so-called trace identity (or variational identity) provide a systematic construction approach for establishing Hamiltonian structures of integrable hierarchies \cite{tu-jmp,tu-jpa,tah,gz}.

WKI type integrable systems \cite{wki, xu-pla, zysm} not only represent the classical WKI equation \cite{wki},
\begin{eqnarray*}
	u_{t}=\left(\frac{u}{\sqrt{1+u^2}}\right)_{xx},
\end{eqnarray*}
but also represent a large class of related equations such as the following short pulse (SP) equation,
\begin{eqnarray*}
	u_{xt}=u+\frac{1}{6}(u^3)_{xx}.
\end{eqnarray*}
SP equation is proposed as a model to describe ultra-short optical pulses traversing within a nonlinear
media \cite{sw, cjsw, sfo}.

Recently, how to use some specific algebra as a tool to systematically construct integrable systems began to develop, which aroused great interest of scholars. In Refs.\cite{gs, fgss, af}, the algebraic construction  based
on Toda field theory is generalized by the addition of
fields associated to higher grading operators, yielding the generalized affine Toda models.
The higher grade fields are physically interpreted as matter fields with the usual Toda
fields coupled to them. In Ref.\cite{fgz}, authors propose a general higher grading construction for the zero curvature
equation, containing the WKI hierarchy as a particular case. In the construction, the
zero grade Toda fields are completely removed, remaining the higher grade fields only.
Based on this method, one can obtain a series of mixed integrable systems such as the mixed mKdV-sine-Gordon equation, the mixed AKNS-Lund-Regge equation and the mixed super symmetric mKdV-sinh-Gordon equation.
Integrable systems obtained from negative flows have important physical and mathematical significance, such as the Camassa-Holm (CH) equation, the Degasperis-Procesi (DP) equation and the Vakhnenko equation. A mixed WKI-SP model has been found by combining the positive flows and extend the WKI hierarchy to incorporate negative flows \cite{fgz, ggz}.

The remainder of this paper is organized as follows.  In section 2,  we would like to construct a standard-form WKI type integrable  hierarchy and bi-Hamiltonian structure by using the  trace identity. In section 3,  we introduce the higher grading affine algebraic construction method and  some special cases in the obtained  hierarchy are considered. The local and non-local conserved charges are obtained from the Riccati form.
In section 4,  higher order SP equation and mixed WKI-SP equation are derived by considering the negative flow and mixed  flow.
  The last section is devoted to conclusions and discussions.

\section{A standard-form WKI type integrable hierarchy}

For the sake of readability, let's first introduce the three-dimensional real special linear Lie algebra $sl(2, \mathbb{R})$.
This algebra consisting
of trace-free $2 \times 2$ matrices,  has the basis
\begin{eqnarray*}
& & e_1=\left(\begin{array} {cc}
		1 &  0  \cr
		0 & -1
\end{array}\right),\qquad e_2=\left(\begin{array} {cc}
		0 & 1 \cr
		0 & 0
\end{array}\right),\qquad e_3=\left(\begin{array} {cccc}
		0 & 0 \cr
		1 & 0
\end{array}\right),
\end{eqnarray*}
whose nonzero commutator relations are
\begin{eqnarray*}
	& &  [e_1, e_2]=2e_2, \qquad [e_1, e_3]=-2e_3, \qquad [e_2, e_3]=e_1.
\end{eqnarray*}
We can also define the corresponding matrix loop algebra $\widetilde{sl}(2, \mathbb{R})$,
\begin{eqnarray*}
	\widetilde{sl}(2, \mathbb{R})=\left\{\sum_{j\geq 0}M_j\lambda^{n-j}\Big|M_j\in {sl}(2, \mathbb{R}), ~ j\geq 0, ~ n\in \mathbb{Z}\right\}.
\end{eqnarray*}
Thus, a brief account of the procedure for building integrable hierarchies associated with $sl(2, \mathbb{R})$ is described below.

\noindent{\it Step 1:}    One needs to select an appropriate spectral matrix $U$ to form  a spatial spectral problem $\phi_x=U\phi$.

\noindent{\it Step 2:}     Construct a particular Laurent series solution $W=W(u, \lambda)$, to the stationary zero curvature equation  $W_x=[U, W]$, based on which one can also prove the localness property for $W$.

\noindent{\it Step 3:} By means of the solution $W$  obtained in the above step, introduce temporal spectral problems $\phi_{t_m}=V^{[m]}\phi$
 so that the zero curvature equations $ U_{t_m}-V_x^{[m]}+\left[U, V^{[m]}\right]=0$ will generate an integrable hierarchy $u_{t_m}=K_m(u), ~ m\geq 0$.

\noindent{\it Step 4:}     Finally, furnish Hamiltonian structures $ u_{t_m}=K_m(u)=J\frac{\delta \mathcal{H}_m}{\delta u}, ~ m\geq 0$ by trace identity.

In Ref.\cite{xu-pla}, a WKI type spatial spectral problem, which is associated with ${sl}(2, \mathbb{R})$, is defined by
\begin{eqnarray}\label{a1}
& & \phi_x=U\phi=U(u,\lambda)\phi,\qquad u=\left(\begin{array} {c}
p \cr
q
\end{array}\right), \qquad \phi=\left(\begin{array} {c}
\phi_1 \cr
\phi_2
\end{array}\right),
\end{eqnarray}
where
\begin{eqnarray}\label{a2}
& & U=(\lambda+\alpha p) e_1+\lambda p e_2+\lambda q e_3\nonumber\\
& &\makebox[3.5mm]{} =\left(\begin{array} {cc}
\lambda+\alpha p & \lambda p \cr
\lambda q & -\lambda-\alpha p
\end{array}\right)\in \widetilde{sl}(2, \mathbb{R}).
\end{eqnarray}
When $\alpha=0$, it is exactly the classical WKI spatial spectral problem \cite{wki}.
In order to construct recursion relations, we consider the following standard-form  of matrix $W$,
\begin{eqnarray}\label{a3}
& & W=a e_1+b e_2+c e_3\nonumber\\
& & \makebox[4.5mm]{} =\left(\begin{array} {cc}
		a &  b \cr
		c & -a
\end{array}\right)\in \widetilde{sl}(2, \mathbb{R}).
\end{eqnarray}

 \begin{remark}
In Ref.\cite{xu-pla}, Matrix $W$ is taken as follows
 \begin{eqnarray*}
& & W=\left(\begin{array} {cc}
		a\lambda +\alpha p a+\frac{\alpha  b_x}{\lambda} &  ap\lambda+b_x \cr
		aq\lambda+c_x & -a\lambda -\alpha p a-\frac{\alpha  b_x}{\lambda}
\end{array}\right).
\end{eqnarray*}
Eq.\eqref{a3} is  easy to show the  correspondence between the Tu scheme method and the higher grading structure construction method.

\end{remark}

Firstly, we  solve the stationary zero curvature equation
\begin{eqnarray*}
	W_{x}=[U,W],
\end{eqnarray*}
which becomes
\begin{equation}\label{a4}
\left\{\begin{array}{l} a_{x}=\lambda (pc-qb),\vspace{2mm}\\
 b_{x}=2\lambda b-2\lambda pa+2\alpha pb,\vspace{2mm}\\
  	c_{x}=-2\lambda c+2\lambda qa-2\alpha pc.\end{array}\right.
\end{equation}
Substituting  the following Laurent series expansion
\begin{eqnarray}\label{a5}
& & \qquad W=\sum_{k=0}^{\infty}W_k\lambda^{-k},\qquad W_k=\left[\begin{array} {cc}
a_k   & b_k \cr
c_k & -a_k
\end{array}\right],\qquad k\ge 0
\end{eqnarray}
 into \eqref{a4}, we have
\begin{equation}\label{2.6}
\left\{ \begin{array}{l}
a_{kx}=pc_{k+1}-qb_{k+1},\vspace{2mm}\\ b_{k+1}=\frac{1}{2}\left(b_{kx}+2pa_{k+1}-2\alpha pb_k\right),\vspace{2mm}
\\ c_{k+1}=\frac{1}{2}\left(-c_{kx}+2qa_{k+1}-2\alpha pc_k\right),\end{array}
\right. \qquad k\geq 0,
\end{equation}
and
\begin{eqnarray}\label{2.7}
& & pc_0-qb_0=0,\qquad b_0=pa_0, \qquad c_0=qa_0.
\end{eqnarray}
To determine the recursion relation between $\left\{b_{k+1}, c_{k+1} \right\}$ and $\left\{b_{k}, c_{k} \right\}$, we need to represent $a_{k+1}$ by $\left\{b_{k}, c_{k} \right\}$.
In order to achieve this purpose, we rewrite $a_{kx}$ as
\begin{eqnarray}\label{2.8}
& & a_{kx}=pc_{k+1}-qb_{k+1}\nonumber\\
& & \qquad=p\left[\frac{1}{2}\left(-c_{kx}+2qa_{k+1}-2\alpha pc_k\right)\right]-q\left[\frac{1}{2}\left(b_{kx}+2pa_{k+1}-2\alpha pb_k\right)\right]\nonumber\\
& & \qquad=-\frac{1}{2}pc_{kx}-\frac{1}{2}qb_{kx}-\alpha p^{2}c_k+\alpha pqb_k.
\end{eqnarray}
Thus we change $a_{k+1,x}$ to
\begin{eqnarray*}
& & a_{k+1,x}=-\frac{1}{2}pc_{k+1,x}-\frac{1}{2}qb_{k+1,x}-\alpha p^{2}c_{k+1}+\alpha pqb_{k+1}\nonumber\\
& & \qquad\quad=-\frac{1}{2}p\left\{\frac{1}{2}
\bigl[-c_{kxx}+2q_xa_{k+1}+2qa_{k+1,x}-2(\alpha pc_k)_x\bigr]\right\}\nonumber\\
& & \qquad\qquad-\frac{1}{2}q\left\{\frac{1}{2}\bigl[b_{kxx}
+2p_xa_{k+1}+2pa_{k+1,x}-2(\alpha pb_k)_x\bigr]\right\}\nonumber\\
& & \qquad\qquad-\alpha p^{2}\left[\frac{1}{2}\left(-c_{kx}+2qa_{k+1}-2\alpha pc_k\right)\right]\nonumber\\
& & \qquad\qquad+\alpha pq\left[\frac{1}{2}\left(b_{kx}+2pa_{k+1}-2\alpha pb_k\right)\right].
\end{eqnarray*}
Let's rewrite the above equation again as
\begin{eqnarray*}
& & \left(\sqrt{pq+1}a_{k+1}\right)_x=\frac{1}{\sqrt{pq+1}}\left[\frac{1}{4}pc_{kxx}-\frac{1}{4}qb_{kxx}+\frac{1}{2}\alpha p(pc_k)_x+\frac{1}{2}\alpha q(pb_k)_x\right.\nonumber\\
& & \qquad\qquad\qquad\qquad \qquad\qquad\quad \left.+\frac{1}{2}\alpha p^{2}c_{kx}+\frac{1}{2}\alpha pqb_{kx}+\alpha^{2}p^{3}c_{k}-\alpha^{2}p^{2}qb_{k}\right].
\end{eqnarray*}
It means that we arrive at
\begin{eqnarray}\label{2.9}
& & a_{k+1}=\frac{1}{\sqrt{pq+1}}\partial^{-1}\frac{1}{\sqrt{pq+1}}\left[\frac{1}{4}pc_{kxx}-\frac{1}{4}qb_{kxx}+\frac{1}{2}\alpha p(pc_k)_x+\frac{1}{2}\alpha q(pb_k)_x\right.\nonumber\\
& & \qquad\qquad\qquad\qquad \qquad\qquad\quad \left.+\frac{1}{2}\alpha p^{2}c_{kx}+\frac{1}{2}\alpha pqb_{kx}+\alpha^{2}p^{3}c_{k}-\alpha^{2}p^{2}qb_{k}\right].
\end{eqnarray}
So  we can compute
$\left\{a_k, b_k, c_k, ~ k\geq 1\right\}$ recursively from the following  initial values
\begin{eqnarray}\label{2.10}
& & a_0=\frac{1}{\sqrt{pq+1}}, \qquad b_0=\frac{p}{\sqrt{pq+1}}, \qquad c_0=\frac{q}{\sqrt{pq+1}},
\end{eqnarray}
by using  Eq.\eqref{2.9} and the last two equations of \eqref{2.6}.
Here $\left\{a_0, b_0, c_0\right\}$ are determined by the initial conditions \eqref{2.7}.
 To guarantee the uniqueness of $\left\{a_k, b_k, c_k\right\}$, we also need impose the integration conditions
\begin{eqnarray*}
& & a_k|_{u=0}=b_k|_{u=0}=c_k|_{u=0}=0, \qquad k\geq 1.
\end{eqnarray*}
Here we  use Maple software
to deal with complicated symbolic computations.
The first two sets are listed as follows:
\begin{eqnarray*}
& & a_1=\frac{1}{4}\frac{pq_{x}-qp_{x}}{(pq+1)^{3/2}}+\alpha\frac{ p^{2}q}{(pq+1)^{3/2}},\\
& & b_1=\frac{1}{2}\frac{p_x}{(pq+1)^{3/2}}-\alpha\frac{p^{2}}{(pq+1)^{3/2}},\\
& & c_1=-\frac{1}{2}\frac{q_x}{(pq+1)^{3/2}}-\alpha\frac{pq}{(pq+1)^{3/2}};\\
& & a_2=-\frac{1}{32}\frac{1}{(pq+1)^{7/2}}\left(4p^2qq_{xx}-5p^2q_x^2-14pqp_xq_x+4pq^2p_{xx}-5q^2p_x^2+4pq_{xx}\right.\\
& & \qquad\quad \left.-4p_xq_x+4qp_{xx}\right)-\frac{3}{4}\alpha\frac{p(pq_x-p_xq)}{(pq+1)^{5/2}}-\frac{3}{2}\alpha^2\frac{p^3q}{(pq+1)^{5/2}},\\
& & b_2=-\frac{1}{32}\frac{1}{(pq+1)^{7/2}}\left(-4p^2q^2p_{xx}+4p^3qq_{xx}-5p^3q_x^2-2p^2qp_xq_x+7pq^2p_x^2-12pqp_{xx}\right.\\
& & \qquad\quad \left.+4p^2q_{xx}+8pp_xq_x+12qp_x^2-8p_{xx}\right)-\frac{3}{2}\alpha\frac{pp_x}{(pq+1)^{5/2}}-\frac{1}{2}\alpha^2\frac{p^3(pq-2)}{(pq+1)^{5/2}},\\
& & c_2=\frac{1}{32}\frac{1}{(pq+1)^{7/2}}\left(-4pq^3p_{xx}+5q^3p_x^2+4p^2q^2q_{xx}+2pq^2p_xq_x-7p^2qq_x^2-4q^2p_{xx}\right.\\
& & \qquad\quad \left.+12pqq_{xx}-8qp_xq_x-12pq_x^2+8q_{xx}\right)
-\frac{1}{2}\alpha\frac{p^2qq_x-pq^2p_x-2pq_x-qp_x}{(pq+1)^{5/2}}\\&&\qquad-\frac{1}{2}\alpha^2\frac{p^2q(pq-2)}{(pq+1)^{5/2}}.
\end{eqnarray*}
 In fact, values of functions  $\left\{a_k, b_k, c_k,k\ge0\right\}$ are all local. We can prove this fact from the recursion relations of the last two equations in Eq.\eqref{2.6} and
 \begin{eqnarray*}
	& & a_{k+1}=-\frac{1}{2\sqrt{pq+1}}\sum_{i+j=k+1 \atop i,j\geq 1}(a_ia_j+b_ic_j)+\frac{pc_{kx}-qb_{kx}}{4(pq+1)}+\alpha\frac{p(pc_k+qb_k)}{2(pq+1)},
\end{eqnarray*}
which is derived from $a^2+bc=\left(a^2+bc\right)\big|_{u=0}=1$.

Now, taking
\begin{eqnarray*}
& & V^{[m]}=\lambda^2(\lambda^{m}W)_++\Delta_m\\
& & \qquad=\lambda^{m+2}W_0+\lambda^{m+1}W_1+\cdots+\lambda^2W_m+\left[\begin{array} {cc}
h_m & \lambda f_m \cr
\lambda g_m & -h_m
\end{array}\right]\\
& & \qquad=\left[\begin{array} {cc}
\sum_{k=0}^ma_k\lambda^{m+2-k}+h_m & \sum_{k=0}^mb_k\lambda^{m+2-k}+\lambda f_m \cr
\sum_{k=0}^mc_k\lambda^{m+2-k}+\lambda g_m & -\sum_{k=0}^ma_k\lambda^{m+2-k}-h_m
\end{array}\right],
\end{eqnarray*}
the zero curvature equations
\begin{eqnarray*}
& & U_{t_m}-V_x^{[m]}+\left[U,V^{[m]}\right]=0, \qquad n\geq 0
\end{eqnarray*}
give
\begin{eqnarray}\label{2.11}
& & f_m=\frac{1}{2}\left(b_{mx}-2\alpha pb_m\right),\nonumber\\
& & g_m=-\frac{1}{2}\left(c_{mx}+2\alpha pc_m\right),\nonumber\\
& & a_{mx}=pg_m-qf_m,\nonumber\\
& & p_{t_m}=f_{mx}-2\alpha pf_m+2ph_m,\nonumber\\
& & q_{t_m}=g_{mx}-2qh_m+2\alpha pg_m,\nonumber\\
& & h_{mx}=\alpha p_{t_m}.
\end{eqnarray}
Due to (\ref{2.6}), we can easily see that
\begin{eqnarray*}
& &f_{m}=\frac{1}{2}(b_{mx}-2\alpha pb_{m})=b_{m+1}-pa_{m+1},\nonumber\\
& &g_{m}=-\frac{1}{2}(c_{mx}+2\alpha pc_{m})=c_{m+1}-qa_{m+1},\\
& &a_{mx}=pc_{m+1}-qb_{m+1},\nonumber
\end{eqnarray*}
and
\begin{eqnarray*}
\begin{aligned}
		h_{mx}&=\alpha p_{t_{m}}=\alpha(f_{mx}-2\alpha pf_{m}+2ph_{m})\\
		&=\alpha[b_{m+1,x}-(pa_{m+1})_{x}-2\alpha p(b_{m+1}-pa_{m+1})+2ph_{m}].
\end{aligned}
\end{eqnarray*}
Thus we can introduce
\begin{eqnarray*}
	h_{m}=-\alpha pa_{m+1}+\alpha b_{m+1},
\end{eqnarray*}
and then  $p_{t_{m}}$, $q_{t_{m}}$ can be expressed as follows
\begin{eqnarray*}
&&p_{t_{m}}=f_{mx}-2\alpha pf_{m}+2ph_{m}\\
&&\qquad =\frac{1}{2}b_{mxx}-\alpha(pb_{m})_{x}-\alpha p(b_{mx}-2\alpha pb_{m})+2p(-\alpha pa_{m+1}+\alpha b_{m+1})\\
&&\qquad=\frac{1}{2}b_{mxx}-\alpha(pb_{m})_{x},\\
&&	q_{t_{m}}=-\frac{1}{2}c_{mxx}-\alpha(pc_{m})_{x}+2\alpha a_{mx}.
\end{eqnarray*}	
 Therefore, we have obtained a WKI type integrable hierarchy associated with the Lie algebra ${sl}(2,\mathbb{R})$:
\begin{eqnarray}\label{a12}
& & \left[\begin{array} {c}
p \cr
q
\end{array}\right]_{t_m}=K_m=\left[\begin{array} {c}
\frac{1}{2}b_{mxx}-\alpha(pb_{m})_{x} \cr
-\frac{1}{2}c_{mxx}-\alpha(pc_{m})_{x}+2\alpha a_{mx}
\end{array}\right].
\end{eqnarray}
When $\alpha=0$, it just is the classical WKI integrable hierarchy \cite{wki}.

Next, we construct Hamiltonian structures of the above WKI type integrable hierarchy \eqref{a12},  which are furnished by using the
following trace identity \cite{tu-jmp,tu-jpa,tah,gz},
\begin{eqnarray*}
& & \frac{\delta}{\delta u}\int{\rm tr}\left(\frac{\partial U}{\partial \lambda}W\right){\rm d}x=\lambda^{-\gamma}\frac{\partial}{\partial \lambda}\lambda^{\gamma}{\rm tr}\left(\frac{\partial U}{\partial u}W\right), \qquad \gamma=-\frac{\lambda}{2}\frac{{\rm d}}{{\rm d}\lambda}{\rm ln}\left|{\rm tr}\left(W^2\right)\right|,
\end{eqnarray*}
 where $W$ solves the stationary zero curvature equation $W_x=[U,W]$.
 Thus, the corresponding trace identity becomes
\begin{eqnarray}\label{a14}
& &  \frac{\delta}{\delta u}\int \left(-\frac{2a_{m} +qb_{m}+pc_{m}}{m-1}\right) {\rm d}x=\left(\begin{array} {c}
 c_{m}+2\alpha a_{m-1} \cr
 b_{m}
\end{array}\right).
\end{eqnarray}
By means of Eq.\eqref{a12}, we can compute
\begin{eqnarray}
\begin{aligned}
	p_{t_{m}}&=\frac{1}{2}b_{m,xx}-\alpha(pb_{m})_{x}=\frac{1}{2}\partial^{2}b_{m}-\alpha\partial(pb_{m}),\nonumber\\
	q_{t_{m}}&=-\frac{1}{2}c_{m,xx}-\alpha(pc_{m})_{x}+2\alpha a_{mx}\\               &=-\frac{1}{2}\partial^{2}c_{m}-\alpha\partial(pc_{m})
	+2\alpha(-\frac{1}{2}pc_{mx}-\frac{1}{2}qb_{mx}-\alpha p^2c_{m}+\alpha pqb_{m})\\
	&=-\frac{1}{2}\partial^{2}(c_{m}+2\alpha a_{m-1})+\alpha\partial^{2}a_{m-1}
	-\alpha\partial p(c_{m}+2\alpha a_{m-1})+2\alpha^{2}\partial pa_{m-1}-\alpha p\partial (c_{m+1}+2\alpha a_{m-1})\\&\quad+2\alpha^{2}p\partial a_{m-1}-\alpha q\partial b_{m}-2\alpha^{2}p^{2}(c_{m}+2\alpha a_{m-1})+4\alpha^{3}p^{2}a_{m-1}+2\alpha^{2}pqb_{m}\\
	&=(-\frac{1}{2}\partial^{2}-\alpha\partial p-\alpha p\partial
	-2\alpha^{2}p^{2})(c_{m}+2\alpha a_{m-1})+(-\alpha q\partial+2\alpha^{2}pq)b_{m}
	+\alpha\partial(pc_{m}-qb_{m})\\&\quad+2\alpha^{2}\partial pa_{m-1}+2\alpha^{2}p(pc_{m}-qb_{m})+4\alpha^{3}p^{2}a_{m-1}\\
	&=(-\frac{1}{2}\partial^{2}-\alpha p\partial)(c_{m}+2\alpha a_{m-1})+(-\alpha q\partial-\alpha\partial q)b_{m}.\nonumber
\end{aligned},
\end{eqnarray}
Thus,the above integrable hierarchy \eqref{a12} can be represented as the following Hamiltonian forms
\begin{eqnarray*}
& & u_{t_m}=K_m=J\frac{\delta \mathcal{H}_m}{\delta u},
\end{eqnarray*}
with the Hamiltonian operator
\begin{eqnarray*}
& & J=\left(\begin{array} {cc}
0 & \frac{1}{2}\partial^{2}-\alpha\partial p     \cr
-\frac{1}{2}\partial^{2}-\alpha p\partial &  -\alpha q\partial-\alpha\partial q
\end{array}\right)
\end{eqnarray*}
and the Hamiltonian functionals
\begin{eqnarray*}
& &  \mathcal{H}_m= \int \left(-\frac{2a_{m} +qb_{m}+pc_{m}}{m-1}\right) {\rm d}x, \qquad m\geq 2.
\end{eqnarray*}
It is now a direct computation that all members in Eq.\eqref{a12} are bi-Hamiltonian.
We compute
\begin{eqnarray*}
\begin{aligned}
	c_{m}+2\alpha a_{m-1}&=-\frac{1}{2}c_{m-1,x}-\alpha pc_{m-1}
	+\frac{q}{\sqrt{pq+1}}\partial^{-1} \frac{1}{\sqrt{pq+1}}[\frac{1}{4}pc_{m-1,xx}	-\frac{1}{4}qb_{m-1,xx}\\&\quad+\frac{1}{2}\alpha p(pc_{m-1})_{x}+\frac{1}{2}\alpha q(pb_{m-1})_{x}+\frac{1}{2}\alpha p^{2}c_{m-1,x}+\frac{1}{2}\alpha pqb_{m-1,x}+\alpha^{2}p^{3}c_{m-1}\\&\quad-\alpha^{2}p^{2}qb_{m-1}]+2\alpha\partial^{-1}(-\frac{1}{2}pc_{m-1,x}-\frac{1}{2}qb_{m-1,x}-\alpha p^2c_{m-1}+\alpha pqb_{m-1})\\
	&=\Psi_{11}(c_{m-1}+2\alpha a_{m-2})+\Psi_{12}b_{m-1},
\end{aligned}
\end{eqnarray*}
with
\begin{eqnarray*}
	&&\Psi_{11}=-\frac{1}{2}\partial+\frac{q}{\sqrt{pq+1}}\partial^{-1} \frac{1}{\sqrt{pq+1}}(\frac{1}{4}p\partial^{2}+\frac{1}{2}\alpha p^{2}\partial)-\alpha\partial^{-1}p\partial,\\
	&&\Psi_{12}=\frac{q}{\sqrt{pq+1}}\partial^{-1}\frac{1}{\sqrt{pq+1}}
	(-\frac{1}{4}q\partial^{2}+\frac{1}{2}\alpha q\partial p+\frac{1}{2}\alpha pq\partial+\frac{1}{2}\alpha p\partial q)-\alpha\partial^{-1}q\partial-\alpha q.
\end{eqnarray*}
Similarly, we have
\begin{eqnarray*}
	b_{m}=\Psi_{21}(c_{m-1}+2\alpha a_{m-2})+\Psi_{22}b_{m-1},
\end{eqnarray*}
where
\begin{eqnarray*}
	&&\Psi_{21}=\frac{p}{\sqrt{pq+1}}\partial^{-1}\frac{1}{\sqrt{pq+1}}
	(\frac{1}{4}p\partial^{2}+\frac{1}{2}\alpha p^{2}\partial),\\
	&&\Psi_{22}=\frac{1}{2}\partial+\frac{p}{\sqrt{pq+1}}\partial^{-1} \frac{1}{\sqrt{pq+1}}(-\frac{1}{4}q\partial^{2}+\frac{1}{2}\alpha q\partial p+\frac{1}{2}\alpha pq\partial+\frac{1}{2}\alpha p\partial q)-\alpha p.
\end{eqnarray*}
So we arrive at
\begin{eqnarray}\label{a16}
	U_{t_{m}}=K_{m}=J\frac{\delta H_{m}}{\delta u}=\Psi\frac{\delta H_{m-1}}{\delta u},
\end{eqnarray}
where
\begin{eqnarray*}
\Psi=\left(\begin{array} {cc}
\Psi_{11} & \Psi_{12}    \\
\Psi_{21} &  \Psi_{22}
\end{array}\right).
\end{eqnarray*}
Therefore, it is easy to see that the WKI type integrable hierarchy \eqref{a12} is Liouville integrable.

\section{The higher grading construction}

Firstly, we give a brief description of the procedure for the higher grading construction method. A more detailed description of the method process is given in Ref.\cite{fgz}.
Let $\widehat{\mathcal{G}}$ be an affine Kac-Moody algebra and $Q$ an operator decomposing the algebra into
the graded subspaces
$$\widehat{\mathcal{G}}={\bigoplus \limits_{j\in \mathbb{Z}}}\widehat{\mathcal{G}}^{(j)}, \qquad [Q, ~ \widehat{\mathcal{G}}^{(j)}]= j\widehat{\mathcal{G}}^{(j)}.$$
As a consequence of the Jacobi identity, $[\widehat{\mathcal{G}}^{(i)}, ~ \widehat{\mathcal{G}}^{(j)}]\subset \widehat{\mathcal{G}}^{(i+j)}$.
Let $E$ be a semi-simple element, with a definite grade, defining the kernel subspace
$$\mathcal{K}=\{T\in\widehat{\mathcal{G}}\mid[E, ~ T]=0\}.$$
The image subspace, $\mathcal{M}$, is its complement and  $\widehat{\mathcal{G}}=\mathcal{M}\oplus\mathcal{K}$.
Then we have the relations $[\mathcal{K}, ~ \mathcal{K}]\subset\mathcal{K}$, $[\mathcal{K}, ~ \mathcal{M}]\subset\mathcal{M}$
and we assume the symmetric space structure $[\mathcal{M}, ~ \mathcal{M}]\subset\mathcal{K}$.

Integrable systems can be  constructed from the zero curvature equation
\begin{eqnarray}\label{c1}
	[\partial_{x}+U, ~ \partial_{t}+V]=0,	
\end{eqnarray}	
where $U$ and $V$ lied on $\widehat{\mathcal{G}}$ and have the following forms
\begin{eqnarray}\label{c2}
&&	U=E^{(0)}+E^{(1)}+A^{(1)}[\phi],\nonumber\\
&&	V=\sum_{i=-m}^{n}D^{[i]}[\phi].
\end{eqnarray}	
Here $E^{(0)}\in\widehat{\mathcal{G}}_{0}$ and $E^{(1)}\in\widehat{\mathcal{G}}^{(1)}$ is a constant semi-simple element.  $A^{(1)}[\phi]\in\mathcal{M}^{(1)}$, where
$\mathcal{M}^{(1)}=\mathcal{M}\bigcap \widehat{\mathcal{G}}^{(1)}$ is the operator containing the field functions now having grade one.
Then substituting Eq.\eqref{c2} into Eq.\eqref{c1}, the zero curvature representation becomes
\begin{eqnarray*}\label{c6}
	[\partial_{x}+E^{(0)}+E^{(1)}+A^{(1)}, ~ \partial_{t}+D^{(-m)}+D^{(-m+1)}+\dots+D^{(-1)}+D^{(0)}+\dots+D^{(n-1)}+D^{(n)}]=0.
\end{eqnarray*}
With this algebraic structure, this zero curvature equation can be solved non trivially. The projection into each graded subspace yields
the following set of equations
\begin{eqnarray*}
	[E^{(1)}+A^{(1)},D^{(n)}]&=0,\label{3.7}\\
	\partial_{x}D^{(n)}+[E^{(0)},D^{(n)}]+[E^{(1)}+A^{(1)},D^{(n-1)}]&=0,\label{3.8}\\
	\quad\quad\quad\vdots\nonumber\\
	\partial_{x}D^{(1)}+[E^{(0)},D^{(1)}]+[E^{(1)}+A^{(1)},D^{(0)}]
	-\partial_{t}(E^{(1)}+A^{(1)})&=0,\label{3.9}\\
	\partial_{x}D^{(0)}+[E^{(0)},D^{(0)}]+[E^{(1)}+A^{(1)},D^{(-1)}]
	-\partial_{t}E^{(0)}&=0,\label{3.10}\\
	\quad\quad\quad\vdots\nonumber\\
	\partial_{x}D^{(-m+1)}+[E^{(0)},D^{(-m+1)}]+[E^{(1)}+A^{(1)},D^{(-m)}]&=0,\label{3.11}\\
	\partial_{x}D^{(-m)}+[E^{(0)},D^{(-m)}]&=0.\label{3.12}
\end{eqnarray*}

Now we consider a concrete Kac-Moody algebra  $\hat{A}_{1}=\{H^{j},~ E_{+}^{j}, ~ E_{-}^{j}, ~ \hat{c}, ~ \hat{d}\}$ with commutation relations,
\begin{eqnarray*}
&&[H^{k},~ H^{j}]=2k\delta_{k+j,0}\hat{c},\qquad[H^{k}, ~ E_{\pm}^{j}]=\pm2kE_{\pm}^{k+j},\qquad
[E_{+}^{k}, ~ E_{-}^{j}]=H^{k+j}+k\delta_{k+j,0}\hat{c},\\
&&[\hat{d}, ~ T^{j}]=jT^{j},\qquad\quad\qquad[\hat{c}, ~ T^{j}]=0,	
\end{eqnarray*}
where $T^{j}\in\{H^{j},E_{+}^{j},E_{-}^{j}\}$ and $j$ is an integer.
For construction of integrable systems we just need to use the loop algebra, achieved by setting $\hat{c}=0$.
The homogeneous gradation $Q=\hat{d}$ yields the grading subspaces
\begin{eqnarray*}
\widehat{\mathcal{G}}^{(j)}=\{H^{j},E_{+}^{j},E_{-}^{j}\}.
\end{eqnarray*}	
Fixing the semi-simple element as $E=\alpha pH^{0}+H^{1}$,  we have $\mathcal{K}^{(j)}=\{H^{j}\}$ and $\mathcal{M}^{(j)}=\{E_{+}^{j},E_{-}^{j}\}$.
Thus the operator containing functions $p\equiv p(x,t), q\equiv q(x,t)$ have the form
\begin{eqnarray*}
	A^{(1)}=pE_{+}^{1}+qE_{-}^{1}.	
\end{eqnarray*}
By setting the Lax operator $V$ which is a sum of
elements in the form
\begin{eqnarray*}
	D^{(j)}=a_{j}E_{+}^{j}+b_{j}E_{-}^{j}+c_{j}H^{j},
\end{eqnarray*}
we have the following zero curvature equation of positive flow
\begin{eqnarray}\label{c1}
	[\partial_{x}+\alpha pH^{0}+H^{1}+pE_{+}^{1}+qE_{-}^{1}, ~ \partial_{t}+D^{(n)}+D^{(n-1)}+\dots+D^{(0)}]=0.
\end{eqnarray}
Here coefficients $a_j, b_j, c_j$ will be determined
in terms of the field functions $p$ and $q$.

When $n=2$, the grad-by-grad decomposing of the above equation \eqref{c1} leads to
\begin{eqnarray*}
	[H^{1}+pE_{+}^{1}+qE_{-}^{1},D^{(2)}]&=0,\\
	\partial_{x}D^{(2)}+[\alpha pH^{0},D^{(2)}]+[H^{1}+pE_{+}^{1}+qE_{-}^{1},D^{(1)}]&=0,\\
	\partial_{x}D^{(1)}+[\alpha pH^{0},D^{(1)}]+[H^{1}+pE_{+}^{1}+qE_{-}^{1},D^{(0)}]
	-\partial_{t}(H^{1}+pE_{+}^{1}+qE_{-}^{1})&=0,\\
	\partial_{x}D^{(0)}+[\alpha pH^{0},D^{(0)}]-\partial_{t}(\alpha pH^{0})&=0.
\end{eqnarray*}
Therefore, we can get the following WKI type integrable system,	
\begin{eqnarray}\label{d1}
	&&\partial_{t}p+\partial_{x}^{2}\Bigg(\frac{p}{2(1+pq)^{1/2}}\Bigg)+\partial_{x}\Bigg(\frac{\alpha p^{2}}{(1+pq)^{1/2}}\Bigg)=0,\nonumber\\
	&&\partial_{t}q-\partial_{x}^{2}\Bigg(\frac{q}{2(1+pq)^{1/2}}\Bigg)+\partial_{x}\Bigg(\frac{\alpha (pq-2)}{(1+pq)^{1/2}}\Bigg)=0.	
\end{eqnarray}
whose Lax pair is given by
\begin{eqnarray*}
	&&U=\alpha pH^{0}+H^{1}+pE_{+}^{1}+qE_{-}^{1},\label{3.18}\\	 &&V=\frac{p}{(1+pq)^{1/2}}E_{+}^{2}+\frac{q}{(1+pq)^{1/2}}E_{-}^{2}+\frac{1}{(1+pq)^{1/2}}H^{2}\nonumber\\&&\quad-\Bigg[\partial_{x}\Bigg(\frac{p}{2(1+pq)^{1/2}}\Bigg)+\frac{\alpha p^{2}}{(1+pq)^{1/2}}\Bigg]E_{+}^{1}+\Bigg[\partial_{x}\Bigg(\frac{q}{2(1+pq)^{1/2}}\Bigg)-\frac{\alpha pq}{(1+pq)^{1/2}}\Bigg]E_{-}^{1}\nonumber\\&&\quad
	-\Bigg[\partial_{x}\Bigg(\frac{\alpha p}{2(1+pq)^{1/2}}\Bigg)+\frac{\alpha^{2}p^{2}}{(1+pq)^{1/2}}\Bigg]H^{0}.\label{3.19}
\end{eqnarray*}	
This equation \eqref{d1} just is the first equation in the WKI type integrable hierarchy \eqref{a12} with $m=0$ by replacing $-U, -V$ with $U, V$ respectively.
In fact, we can find that there is a correspondence between the method in section 2 and this method.
In other words, this method gives a Kac-Moody algebraic interpretation of the Tu  scheme method.

Similarly, when $n=3$, we can construct the following WKI type integrable system,
\begin{eqnarray}
	&&\partial _{t}p-\partial_{x}^{2}\Bigg(\frac{p_{x}+2\alpha p^{2}}{4(1+pq)^{3/2}}\Bigg)
	-\partial_{x}\Bigg(\frac{\alpha p(p_{x}+2\alpha p^{2})}{2(1+pq)^{3/2}}\Bigg)=0,\nonumber\\
	&&\partial _{t}q-\partial_{x}^{2}\Bigg(\frac{q_{x}-2\alpha pq}{4(1+pq)^{3/2}}\Bigg)
	+\partial_{x}\Bigg(\frac{\alpha p(q_{x}-2\alpha pq)}{2(1+pq)^{3/2}}\Bigg)
	+2\alpha\partial_{x}\Bigg(\frac{qp_{x}-pq_{x}+4\alpha p^{2}q}{4(1+pq)^{3/2}}\Bigg)=0,\label{d2}	
\end{eqnarray}
with Lax pair
\begin{eqnarray*}
	&&U=\alpha pH^{0}+H^{1}+pE_{+}^{1}+qE_{-}^{1},\label{3.27}\\
	&&V=\frac{p}{(1+pq)^{1/2}}E_{+}^{3}+\frac{q}{(1+pq)^{1/2}}E_{-}^{3}+\frac{1}{(1+pq)^{1/2}}H^{3}\nonumber\\&&\quad\quad-\frac{p_{x}+2\alpha p^{2}}{2(1+pq)^{3/2}}E_{+}^{2}
	+\frac{q_{x}-2\alpha pq}{2(1+pq)^{3/2}}E_{-}^{2}+\frac{qp_{x}-pq_{x}+4\alpha p^{2}q}{4(1+pq)^{3/2}}H^{2}\nonumber\\&&\quad\quad+\Bigg[\partial_{x}\Bigg(\frac{p_{x}+2\alpha p^{2}}{4(1+pq)^{3/2}}\Bigg)+\frac{\alpha p(p_{x}+2\alpha p^{2})}{2(1+pq)^{3/2}}\Bigg]E_{+}^{1}+\Bigg[\partial_{x}\Bigg(\frac{q_{x}-2\alpha pq}{4(1+pq)^{3/2}}\Bigg)-\frac{\alpha p(q_{x}-2\alpha pq)}{2(1+pq)^{3/2}}\Bigg]E_{-}^{1}\nonumber\\&&\quad\quad+\Bigg[\partial_{x}\Bigg(\frac{\alpha (p_{x}+2\alpha p^{2})}{4(1+pq)^{3/2}}\Bigg)+\frac{\alpha^{2} p(p_{x}+2\alpha p^{2})}{2(1+pq)^{3/2}}\Bigg]H^{0}.\label{3.28}
\end{eqnarray*}
This equation \eqref{d2} just is the second equation in the WKI type integrable hierarchy \eqref{a12} with $m=1$ by replacing $-U, -V$ with $U, V$ respectively..

\begin{remark}
If setting $E=\alpha\sqrt{pq+1}H^{0}+H^{1}$ proposed in Ref.\cite{zysm},  we can get similar results according to the above method.
\end{remark}

Next we shall derive local and nonlocal charges from the Riccati form of the spectral problem
\begin{eqnarray}\label{d42}
(\partial_x+U)\Psi=0,\qquad (\partial_t+V)\Psi=0.
\end{eqnarray}
By using the matrix representation, we have
\begin{eqnarray}\label{d43}
U=\left(\begin{array} {cc}
\lambda+\alpha p & \lambda p    \\
       \lambda q &  -\lambda-\alpha p
\end{array}\right),\qquad V=\left(\begin{array} {cc}
A(\lambda) & B(\lambda)\\
		C(\lambda) & -A(\lambda)
\end{array}\right), \qquad \Psi=\left(\begin{array} {c}
\Psi_1 \\
		\Psi_2
\end{array}\right),
\end{eqnarray}
where $\lambda$ is the  spectral parameter.
Introducing the variables
\begin{eqnarray}\label{d44}
	\Gamma=\frac{\Psi_{2}}{\Psi_{1}},\qquad\Gamma^{-1}=\frac{\Psi_{1}}{\Psi_{2}},
\end{eqnarray}
 we can write the Riccati form of the spectral problem \eqref{d42}, whose compatibility yields the conservation laws
\begin{eqnarray}\label{d46}	
	&&\partial_{t}(\lambda p\Gamma+\alpha p)=\partial_{x}(A+B\Gamma),\nonumber\\
	&&\partial_{t}(\lambda q\Gamma^{-1}-\alpha p)=\partial_{x}(-A+C\Gamma^{-1}).
\end{eqnarray}
Therefore, we can construct an infinite number of conserved charges by using $p\Gamma$ and $q\Gamma^{-1}$, assuming a power series in $\lambda$. Let $F=p\Gamma$ and $G=q\Gamma^{-1}$, we can obtain
\begin{eqnarray}\label{d48}
	&&p(\frac{F}{p})_{x}=2(\lambda+\alpha p)F-\lambda pq+\lambda F^{2},\nonumber\\
	&&q(\frac{G}{q})_{x}=-2(\lambda+\alpha p)G-\lambda pq+\lambda G^{2},
\end{eqnarray}
which are the generating equations for the conserved densities.

To get the local density, we expand $F$ in the power series of $1/\lambda$,
\begin{eqnarray}\label{d49}
	F=\sum_{n=0}^{\infty}f_{n}\lambda^{-n}.
\end{eqnarray}
Substituting (\ref{d49}) into (\ref{d48}) and equating the terms of the same powers of $1/\lambda$, we obtain  conserved densities,
\begin{eqnarray*}
	&&f_{0}=-1+(1+pq)^{1/2},\\
	&&f_{1}=-\frac{1}{2}\partial_{x}\Bigg(\ln\frac{p}{(1+pq)^{1/2}}\Bigg)+\frac{1}{2}\frac{\partial_{x}p+2\alpha p^2}{p(1+pq)^{1/2}}-\alpha p,\\
    &&\quad \vdots \nonumber
\end{eqnarray*}	
The charges associated to these densities are	
\begin{eqnarray*}	
	&&H_{0}=\int_{-\infty}^{\infty}\Bigg(-1+(1+pq)^{1/2}\Bigg)dx,\\
	&&H_{1}=\int_{-\infty}^{\infty}\Bigg(\frac{1}{2}\frac{\partial_{x}p+2\alpha p^2}{p(1+pq)^{1/2}}-\alpha p\Bigg)dx,\\
	 &&\quad \vdots \nonumber
\end{eqnarray*}	
These charges are the Hamiltonians generating the WKI type integrable systems within the positive flows.

To consider furthermore, we set the most general expansion,
\begin{eqnarray}\label{d54}
	F=\sum_{n=-1}^{\infty}f_{-n}\lambda^{n},
\end{eqnarray}
 which can  generate nonlocal densities. Substituting Eq.\eqref{d54} into  Eq.\eqref{d48}, we can get the following system
\begin{eqnarray*}
	&&\partial_{x}f_{1}-(\partial_{x}\ln p+2\alpha p)f_{1}=f_{1}^{2},\\
	&&\partial_{x}f_{0}-(\partial_{x}\ln p+2f_{1}+2\alpha p)f_{0}=2f_{1},\\
	&&\partial_{x}f_{-1}-(\partial_{x}\ln p+2f_{1}+2\alpha p)f_{-1}=f_{0}^{2}+2f_{0}-pq,\\
	&&\partial_{x}f_{-2}-(\partial_{x}\ln p+2f_{1}+2\alpha p)f_{-2}=2f_{0}f_{-1}+2f_{-1},\\
	&&\partial_{x}f_{-3}-(\partial_{x}\ln p+2f_{1}+2\alpha p)f_{-3}=f_{-1}^{2}+2f_{0}f_{-2}+2f_{-2}.\\	
	&&\quad\quad\quad\vdots \nonumber
\end{eqnarray*}	
Using the similar method in Ref.\cite{fgz}, we can get
\begin{eqnarray*}
	&&f_{-1}=-P\partial_{x}^{-1}Q,\\
	&&f_{-2}=-2P\partial_{x}^{-2}Q,\\
	&&f_{-3}=-4P\partial_{x}^{-3}Q+P\partial_{x}^{-1}\big(P(\partial_{x}^{-1}Q)^{2}\big).\\
	&&\qquad\vdots	\nonumber
\end{eqnarray*}
Here $P$ and $Q$ are defined as $P=pe^{2\alpha\partial_{x}^{-1}p},~	Q=qe^{-2\alpha\partial_{x}^{-1}p}$.	
The respective conserved charges are give by
\begin{eqnarray}\label{d70}
	H_{-n}=\int_{-\infty}^{\infty}f_{-n}dx, \qquad n=1,2,3\cdots.
\end{eqnarray}
Thus we believe that the charges are conserved, as can be explicitly checked, either from the positive or
negative flows.

\begin{remark}
For the  spectral problem \eqref{d42} with $U=\alpha \sqrt{pq+1}H^{0}+H^{1}+pE_{+}^{1}+qE_{-}^{1}$, the local and nonlocal charges can be worked out in the same way.
\end{remark}

\section{SP type integrable systems}

In this section, we  study SP type integrable systems by using  the Lax operator $V=\sum_{i=-n}^{1}D^{[i]}[\phi]$.
Here we introduce the operator $\partial_{x}^{-1}f(x)=\int_{-\infty}^{x}f(y)dy$ and assume that the fields and its derivatives of any order decay sufficiently fast when $\mid x\mid\to \infty$. Under this condition $\partial_x\partial_{x}^{-1}f(x)=\partial_{x}^{-1}\partial_xf(x)=f(x)$.

According to our construction, the negative flows can be constructed from the zero curvature equation
\begin{eqnarray*}
	[\partial_{x}+H^{1}+pE_{+}^{1}+qE_{-}^{1}, ~ \partial_{t}+D^{(-n)}+D^{(-n+1)}+\dots+D^{(0)}+D^{(1)}]=0.
\end{eqnarray*}
When $n=1$, we can obtain the following two-component SP equation \cite{fgz}
\begin{eqnarray*}
	&&u_{xt}=4u+2\partial_x(uvu_x),\\
    &&v_{xt}=4v+2\partial_x(vuv_x),
\end{eqnarray*}
with corresponding Lax pair
\begin{eqnarray*}
	&&U=H^{1}+u_{x}E_{+}^{1}+v_{x}E_{-}^{1},\\
	&&V=H^{-1}+2uE_{+}^{0}-2vE_{-}^{0}+2uvu_{x}E_{+}^{1}+2uvv_{x}E_{-}^{1}+2uvH^{1}.
\end{eqnarray*}
Contrary to the positive flows of the WKI type equations,
the SP equation does not seem do describe large amplitude solutions.
A multi-component generalization of the above equation with  the same
structure, has also been proposed in Ref.\cite{mats}.
This generalization also can be obtained by considering the untwisted algebra $\hat{A}_{n-1}\sim  \hat{sl}(n)$.
These conclusions have been proposed in Ref.\cite{fgz}, so we continue to consider higher order SP type integrable systems.

When $n=3$, zero curvature representation reads
\begin{eqnarray*}
	[\partial_{x}+H^{1}+pE_{+}^{1}+qE_{-}^{1}, ~ \partial_{t}+D^{(-3)}+D^{(-2)}+D^{(-1)}+D^{(0)}+D^{(1)}]=0,	
\end{eqnarray*}	
which can  decomposes into six independent equations:
\begin{eqnarray*}
	[H^{1}+pE_{+}^{1}+qE_{-}^{1}, ~ D^{(1)}]&=0,\\
	\partial_{x}D^{(1)}+[H^{1}+pE_{+}^{1}+qE_{-}^{1}, ~ D^{(0)}]-\partial_{t}(H^{1}+pE_{+}^{1}+qE_{-}^{1})&=0,\\
	\partial_{x}D^{(0)}+[H^{1}+pE_{+}^{1}+qE_{-}^{1}, ~ D^{(-1)}]&=0,\\	
	\partial_{x}D^{(-1)}+[H^{1}+pE_{+}^{1}+qE_{-}^{1}, ~ D^{(-2)}]&=0,\\	
	\partial_{x}D^{(-2)}+[H^{1}+pE_{+}^{1}+qE_{-}^{1}, ~ D^{(-3)}]&=0,\\	
	\partial_{x}D^{(-3)}&=0.	
\end{eqnarray*}
We solve this system step by step.
The projection into $\widehat{\mathcal{G}}^{(-3)}$ implies that $a_{-3}$, $b_{-3}$ and $c_{-3}$ are all constants. To consider furthermore, if setting $a_{-3}=b_{-3}=0$,  we have $c_{-2}=constant$ and
\begin{eqnarray*}
	a_{-2}=2c_{-3}\partial_{x}^{-1}p,\quad b_{-2}=-2c_{-3}\partial_{x}^{-1}q,	
\end{eqnarray*}
 by calculating the $\widehat{\mathcal{G}}^{(-2)}$ projection.
Then the $\widehat{\mathcal{G}}^{(-1)}$ projection yields
\begin{eqnarray}\label{d4}
	&&a_{-1}=2\partial_{x}^{-1}(pc_{-2}-2c_{-3}\partial_{x}^{-1}p),\nonumber\\
	&&b_{-1}=-2\partial_{x}^{-1}(qc_{-2}+2c_{-3}\partial_{x}^{-1}q),\\
	&&c_{-1}=2c_{-3}\partial_{x}^{-1}(p\partial_{x}^{-1}q+q\partial_{x}^{-1}p).\nonumber	
\end{eqnarray}	
Similarly,we can obtain the following equation from the $\widehat{\mathcal{G}}^{(0)}$ projection
\begin{eqnarray}\label{d5} &&a_{0}=4\partial_{x}^{-1}[pc_{-3}\partial_{x}^{-1}(p\partial_{x}^{-1}q+q\partial_{x}^{-1}p)-\partial_{x}^{-1}(pc_{-2}-2c_{-3}\partial_{x}^{-1}p)],\nonumber\\
 &&b_{0}=-4\partial_{x}^{-1}[qc_{-3}\partial_{x}^{-1}(p\partial_{x}^{-1}q+q\partial_{x}^{-1}p)+\partial_{x}^{-1}(qc_{-2}+2c_{-3}\partial_{x}^{-1}q)],\nonumber\\
	&&c_{0}=2\partial_{x}^{-1}[q\partial_{x}^{-1}(pc_{-2}-2c_{-3}\partial_{x}^{-1}p)+p\partial_{x}^{-1}(qc_{-2}+2c_{-3}\partial_{x}^{-1}q)].
\end{eqnarray}	
The projection into $\widehat{\mathcal{G}}^{(2)}$ implies that $a_{1}=pc_{1}$ and $b_{1}=qc_{1}$. The $\widehat{\mathcal{G}}^{(1)}$ projection yields the field equations plus one constraint
\begin{eqnarray}\label{d6}
	&&\partial_{t}p=\partial_{x}a_{1}+2(a_{0}-pc_{0}),\nonumber\\
	&&\partial_{t}q=\partial_{x}b_{1}+2(qc_{0}-b_{0}),\nonumber	\\
	&&\partial_{x}c_{1}=qa_{0}-pb_{0}.
\end{eqnarray}
Substituting \eqref{d5} into the third equation of \eqref{d6}, and choosing $c_{-2}=0$, we can obtain
\begin{eqnarray*}
 &&c_{1}=\partial_{x}^{-1}(qa_{0}-pb_{0})\\
 &&\makebox[4mm]{}=4c_{-3}\partial_{x}^{-1}[(q\partial_{x}^{-1}p+p\partial_{x}^{-1}q)\partial_{x}^{-1}(q\partial_{x}^{-1}p+p\partial_{x}^{-1}q)]\nonumber\\&&\quad\quad+8c_{-3}\partial_{x}^{-1}(q\partial_{x}^{-3}p+p\partial_{x}^{-3}q).	 \end{eqnarray*}	
Fixing $c_{-3}$, we have  the following nonlocal equations
\begin{eqnarray*} &&\partial_{t}p=\partial_{x}(pc_{1})+2(a_{0}-pc_{0})\\
&&\makebox[5mm]{}=4c_{-3}\partial_{x}p\partial_{x}^{-1}[q\partial_{x}^{-1}p\partial_{x}^{-1}(q\partial_{x}^{-1}p+p\partial_{x}^{-1}q)\\
&&\quad\quad\quad+p\partial_{x}^{-1}q\partial_{x}^{-1}(q\partial_{x}^{-1}p+p\partial_{x}^{-1}q)+2q\partial_{x}^{-3}p+2p\partial_{x}^{-3}q]\\
&&\quad\quad\quad+8c_{-3}[\partial_{x}^{-1}p\partial_{x}^{-1}(q\partial_{x}^{-1}p+p\partial_{x}^{-1}q)+p\partial_{x}^{-1}(q\partial_{x}^{-2}p-p\partial_{x}^{-2}q)+2\partial_{x}^{-3}p],\\
	 &&\partial_{t}q=\partial_{x}(qc_{1})+2(qc_{0}-b_{0})\\
&&\makebox[5mm]{}=4c_{-3}\partial_{x}q\partial_{x}^{-1}[q\partial_{x}^{-1}p\partial_{x}^{-1}(q\partial_{x}^{-1}p+p\partial_{x}^{-1}q)\\
&&\quad\quad\quad+p\partial_{x}^{-1}q\partial_{x}^{-1}(q\partial_{x}^{-1}p+p\partial_{x}^{-1}q)+2q\partial_{x}^{-3}p+2p\partial_{x}^{-3}q]\\
&&\quad\quad\quad+8c_{-3}[\partial_{x}^{-1}q\partial_{x}^{-1}(q\partial_{x}^{-1}p+p\partial_{x}^{-1}q)+q\partial_{x}^{-1}(p\partial_{x}^{-2}q-q\partial_{x}^{-2}p)+2\partial_{x}^{-3}q].	 \end{eqnarray*}	
Introducing a new field function defined through $p=-q=u_{xxx}$, we get the following model
\begin{eqnarray}\label{d7}
	u_{xxxt}=\frac{2}{15}c_{-3}(u_{xx}^{5})_{xx}-16c_{-3}[u_{xxx}(uu_{xx}-\frac{1}{2}u_{x}^{2})]_{x}-\frac{8}{3}c_{-3}u_{xx}^{3}+16c_{-3}u,
\end{eqnarray}
which is the higher order SP equation. Here $c_{-3}$ is an arbitrary constant. The Lax pair of this model reads
\begin{eqnarray*}
	&&U=H^{1}+u_{xxx}E_{+}^{1}-u_{xxx}E_{-}^{1},\\ &&V=H^{-3}+2u_{xx}E_{+}^{-2}+2u_{xx}E_{-}^{-2}-4u_{x}E_{+}^{-1}+4u_{x}E_{-}^{-1}-2u_{xx}^{2}H^{-1}\\
&&\quad\quad+(-\frac{4}{3}u_{xx}^{3}+8u)E_{+}^{0}+(-\frac{4}{3}u_{xx}^{3}+8u)E_{-}^{0}+u_{xxx}(\frac{2}{3}u_{xx}^{4}-16uu_{xx}+8u_{x}^{2})E_{+}^{1}\\
&&\quad\quad-u_{xxx}(\frac{2}{3}u_{xx}^{4}-16uu_{xx}+8u_{x}^{2})E_{-}^{1}+(\frac{2}{3}u_{xx}^{4}-16uu_{xx}+8u_{x}^{2})H^{1},
\end{eqnarray*}
by setting  $c_{-3}=1$ for the convenience of writing.

It is possible to combine a positive flow with a negative flow, so we consider  mixed WKI-SP integrable systems by using the following zero curvature equation
\begin{eqnarray*}
	[\partial_{x}+H^{1}+pE_{+}^{1}+qE_{-}^{1}, ~ \partial_{t}+D^{(3)}+D^{(2)}+D^{(1)}+D^{(0)}+D^{(-1)}+D^{(-2)}+D^{(-3)}]=0.
\end{eqnarray*}
It can  decompose into eight independent equations,
\begin{eqnarray*}
	[H^{1}+pE_{+}^{1}+qE_{-}^{1}, ~ D^{(3)}]&=0,\\
	\partial_{x}D^{(3)}+[H^{1}+pE_{+}^{1}+qE_{-}^{1}, ~ D^{(2)}]&=0,\\
	\partial_{x}D^{(2)}+[H^{1}+pE_{+}^{1}+qE_{-}^{1}, ~ D^{(1)}]&=0,\\
	\partial_{x}D^{(1)}+[H^{1}+pE_{+}^{1}+qE_{-}^{1}, ~ D^{(0)}]-\partial_{t}(H^{1}+pE_{+}^{1}+qE_{-}^{1})&=0,\\
	\partial_{x}D^{(0)}+[H^{1}+pE_{+}^{1}+qE_{-}^{1}, ~ D^{(-1)}]&=0,\\	
	\partial_{x}D^{(-1)}+[H^{1}+pE_{+}^{1}+qE_{-}^{1}, ~ D^{(-2)}]&=0,\\	
	\partial_{x}D^{(-2)}+[H^{1}+pE_{+}^{1}+qE_{-}^{1}, ~ D^{(-3)}]&=0,\\	
	\partial_{x}D^{(-3)}&=0.
\end{eqnarray*}
We can exactly solve each grade projection starting from highest to lowest.  From the $\widehat{\mathcal{G}}^{(0)}$ to the $\widehat{\mathcal{G}}^{(4)}$ projection, the process of operation is almost similar to the positive flows in the case $n=3$. Thus, in the $\widehat{\mathcal{G}}^{(0)}$ projection, we have
\begin{eqnarray}\label{4.99}
	&&\partial_{t}p=\partial_{x}a_{1}+2(a_{0}-pc_{0}),\nonumber\\
	&&\partial_{t}q=\partial_{x}b_{1}+2(qc_{0}-b_{0}).	
\end{eqnarray}	
From the $\widehat{\mathcal{G}}^{(4)}$ projection, we have $a_{3}=pc_{3}$ and $b_{3}=qc_{3}$. The $\widehat{\mathcal{G}}^{(3)}$ projection gives
\begin{eqnarray}\label{4.10}
	&&a_{2}=-\frac{1}{2}\partial_{x}a_{3}+pc_{2},\qquad b_{2}=\frac{1}{2}\partial_{x}b_{3}+qc_{2}.
\end{eqnarray}
 The $\widehat{\mathcal{G}}^{(2)}$ projection gives
\begin{eqnarray*}
	&&a_{1}=-\frac{1}{2}\partial_{x}a_{2}+pc_{1},\\
	&&b_{1}=\frac{1}{2}\partial_{x}b_{2}+qc_{1},\\
	&&\partial_{x}c_{1}=qa_{0}-pb_{0}.
\end{eqnarray*}
So we can obtain
\begin{eqnarray*}
	&&a_{2}=-\frac{1}{2}\Bigg(\frac{p_{x}}{(1+pq)^{3/2}}\Bigg),\qquad b_{2}=\frac{1}{2}\Bigg(\frac{q_{x}}{(1+pq)^{3/2}}\Bigg), \qquad c_{2}=\frac{1}{4}\Bigg(\frac{qp_{x}-pq_{x}}{(1+pq)^{3/2}}\Bigg).
\end{eqnarray*}	
Now we  let some coefficients, that were previously considered constants,
to depend on time $t$, thus providing the non-autonomous ingredient.
For individual flows those coefficients were not interesting because they come as a global factor in the final
equation. From the $\widehat{\mathcal{G}}^{(-3)}$  to $\widehat{\mathcal{G}}^{(-1)}$ projection,  the results are the same as \eqref{d7}.
Therefore,  we get
\begin{eqnarray}\label{h1}
 &&u_{xxxt}=\frac{a(t)}{4}\Bigg(\frac{u_{xxxx}}{(1-u_{xxx}^{2})^{3/2}}\Bigg)_{xx}+\frac{2}{15}b(t)(u_{xx}^{5})_{xx}-16b(t)\Bigg(u_{xxx}(uu_{xx}-\frac{1}{2}u_{x}^{2})\Bigg)_{x}\nonumber\\&&\quad\quad\quad-\frac{8}{3}b(t)u_{xx}^{3}+16b(t)u,
\end{eqnarray}
after choosing $p=-q=u_{xxx}$. Here $a(t)$ and $b(t)$ are arbitrary functions. This equation \eqref{h1} just is a higher order mixed WKI-SP integrable system.
Due to $a(t)$ and $b(t)$ the dispersion relation will
have a time depend velocity and the solitons will accelerate \cite{fgz}. Eq.\eqref{h1} may be nice
candidates in applications having accelerated ultra-short optical pulses\cite{fgz}.

\section{Conclusions and discussions}

In this paper, we have constructed a standard-form  WKI type integrable hierarchy \eqref{a12}, together with a bi-Hamiltonian structure \eqref{a16} by using the trace identity. This method can be used to other integrable hierarchies as well.

By using the  higher grading construction method, we  give a Kac-Moody  algebraic interpretation of some special equations in this hierarchy \eqref{a12}. In fact, the higher grading construction method is general and other affine Lie
algebras can be considered. we might be able to  use this method to construct novel integrable models.
We also have derived local and nonlocal charges from the Riccati form of the spectral problem \eqref{d42}.

We have extended this WKI type integrable hierarchy to negative flow, which yields a higher order SP type integrable system \eqref{d7}.
A novel integrable non-autonomous WKI-SP equation \eqref{h1} is also proposed, mixing a positive with a negative flow. This mixed model may have applications in nonlinear optics, specially concerning accelerated ultra-short optical pulses \cite{fgz}.

Short pulses and their properties are a subject of current interest in nonlinear optics
and electrodynamics, both theoretically and experimentally \cite{hnw}. In Refs.\cite{sfo, mats, mats2}, authors develop a systematic procedure for constructing exact solutions of SP type equations based the hodograph transformation and the KP reduction technology.
 We can construct the following multi-component SP type system
\begin{eqnarray}\label{p1}
{q_i}_{xt}=q_i+\left[\left(\sum_{j=1}^N\sigma_j|q_j|^2\right){q_i}_x\right]_x-\left(\sum_{j=1}^N\sigma_j|{q_j}_x|^2\right)q_i, \qquad i=1,2,\cdots, N.
\end{eqnarray}
Specifically, when $n=2$ and $\sigma_1=\sigma_2=1$, we have
\begin{eqnarray}\label{p2}
& &   {q_1}_{xt}=q_1+\left[\left(|q_1|^2+|q_2|^2\right){q_1}_x\right]_x-\left(|{q_1}_x|^2+|{q_2}_x|^2\right)q_1,\nonumber\\
& &   {q_2}_{xt}=q_2+\left[\left(|q_1|^2+|q_2|^2\right){q_2}_x\right]_x-\left(|{q_1}_x|^2+|{q_2}_x|^2\right)q_2,
\end{eqnarray}
with the $4\times 4$ Lax pair
\begin{eqnarray*}
& &U=\lambda\left(\begin{array} {cc}
(1-|{q_1}_x|^2-|{q_2}_x|^2)I_2 & 2 \widetilde{q}_x    \\
     2 \widehat{q}_x  &  -(1-|{q_1}_x|^2-|{q_2}_x|^2)I_2
\end{array}\right),\nonumber\\
& & V=\left(\begin{array} {cc}
\lambda(|q_1|^2+|q_2|^2)(1-|{q_1}_x|^2-|{q_2}_x|^2)I_2+\frac{1}{4\lambda}I_2 & 2\lambda(|q_1|^2+|q_2|^2)\widetilde{q}_x-\widetilde{q} \\
		2\lambda(|q_1|^2+|q_2|^2)\widehat{q}_x-\widehat{q}  & -\lambda(|q_1|^2+|q_2|^2)(1-|{q_1}_x|^2-|{q_2}_x|^2)I_2-\frac{1}{4\lambda}I_2
\end{array}\right).
\end{eqnarray*}
Here $I_2$ is the $2\times 2$  identity matrix and $\widetilde{q}= \left(\begin{array} {cc}
q_1 & q_2   \\
   -q_2^*  & q_1^*
\end{array}\right)$, $\widehat{q}= \left(\begin{array} {cc}
q_1^* & -q_2   \\
   q_2^*  & q_1
\end{array}\right)$.
The exact solutions and physical applications of these equations need to be further studied.

\section*{Acknowledgements}

This work is in part supported by
the national natural science foundation of China (Grant No. 11771395).

\vspace{0.3cm}


\begin{thebibliography}{sl}






\bibitem{akns} M.J. Ablowitz, D.J. Kaup, A.C. Newell, H. Segur, The inverse scattering transform-Fourier analysis for nonlinear problems, Stud. Appl. Math. 53 (1974) 249-315.
    
    
\bibitem{kn} D.J. Kaup, A.C. Newell, An exact solution for a derivative nonlinear Schr\"{o}inger equation, J. Math. Phys. 19 (1978) 798-801.

\bibitem{wki} M. Wadati, K. Konno, Y.H. Ichikawa, New integrable nonlinear evolution equations, J. Phys. Soc. Jpn. 47 (1979) 1698-1700.


\bibitem{ff} A.S. Fokas, B. Fuchssteiner,  The hierarchy of the Benjamin-Ono equation, Phys. Lett. A 86 (1981) 341-345.



\bibitem{aks} T.M. Alberty, T. Koikawa, R. Sasaki. Canonical structure of soliton equations I, Physica D 5 (1982) 43-65.

\bibitem{bpt}  M. Boiti, F. Pempinelli, G.Z. Tu, The nonlinear evolution equations related to the Wadati-Konno-Ichikawa spectral problem, Prog. Theor. Phys. 69 (1983) 48-64.




\bibitem{af1}  M. Antonowicz, A.P. Fordy, Coupled KdV equations with multi-Hamiltonian structures. Physica D 28 (1987) 345-357.




\bibitem{af2}   M. Antonowicz, A.P. Fordy, Coupled Harry Dym equations with multi-Hamiltonian structures, J. Phys. A: Math. Gen. 21 (1988) L269-275.







\bibitem{tu-jmp} G.Z. Tu, The trace identity, a powerful tool for constructing the Hamiltonian structure of integrable systems, J. Math. Phys. 30 (1989) 330-338.

\bibitem{tu-jpa} G.Z. Tu, A trace identity and its applications to the theory of discrete integrable systems, J. Phys. A: Math. Gen. 23 (1990) 3903-3922.




\bibitem{tah} G.Z. Tu, R.I. Andrushkiw, X.C. Huang, A trace identity and its application to integrable systems of $1+2$ dimensions, J. Math. Phys. 32 (1991) 1900-1907.

\bibitem{gz}  F.K. Guo, Y.F. Zhang, The quadratic-form identity for constructing the Hamiltonian structure of integrable systems, J. Phys. A: Math. Gen. 38 (2005) 8537-8548.




  \bibitem{xu-pla} X.X. Xu, A generalized Wadati-Konno-Ichikawa hierarchy and new finite-dimensional integrable systems, Phys. Lett. A 301 (2002) 250-262.


\bibitem{zysm} H.Y. Zhu, S.M. Yu, S.F. Shen, Wen-Xiu Ma, New integrable sl(2,R)-generalization of the classical Wadati-Konno-Ichikawa hierarchy, Commun. Nonlinear Sci. Numer. Simulat. 22 (2015) 1341-1349.

\bibitem{sw} T. Sc$\ddot{a}$fer, C.E. Wayne, Propagation of ultra-short optical pulses in cubic nonlinear media, Physica D 196 (2004) 90-105.

\bibitem{cjsw} Y. Chung, C.K.R.T. Jones, T. Sc$\ddot{a}$fer, C.E. Wayne, Ultra-short pulses in linear and nonlinear media, Nonlinearity 18 (2005) 1351-1374.

\bibitem{sfo} S.F. Shen, B.F. Feng, Y. Ohta, From the real and complex coupled dispersionless equations to the real and complex short pulse equations, Stud. Appl. Math. 136 (2016) 64-88.









\bibitem{gs} J.-L. Gervais, M.V. Saveliev, Higher grading generalizations of the Toda
systems, Nucl. Phys. B 453 (1995) 449-476.



\bibitem{fgss} L.A. Ferreira, J.-L. Gervais, J. Sanchez Guillen,M. Saveliev, Affine Toda
systems coupled to matter fields, Nucl. Phys. B 470 (1996) 236-290.




\bibitem{af} P. Assis, L. Ferreira, The Bullough-Dodd model coupled to matter fields, Nucl.
Phys. B 800 (2008) 409-449.





\bibitem{fgz} G.S. Franca, J.F. Gomes, A.H. Zimerman, The higher grading structure of the WKI hierarchy and the two-component short pulse equation, J. High Energ. Phys. 08 (2012) 120.




\bibitem{ggz}  X.G.Geng, F.Y.Guo, Y.Y.Zhai, Two integrable generalizations of WKI and FL equations: Positive and negative flows, and conservation laws, Chinese Phys. B  29 (2020) 70-73.


\bibitem{mats} Y. Matsuno, A novel multi-component generalization of the short pulse equation and its multisoliton solutions, J. Math. Phys. 52 (2011) 123702.


\bibitem{lm}  H. Leblond, D. Mihalache, Few-optical-cycle solitons: Modified Korteweg-de
Vries sine-Gordon equation versus other nonlinear
slowly-varying-envelope-approximation models, Phys. Rev. A 79 (2009) 063835.



\bibitem{hnw}  A.N.W. Hone, V. Novikov, J.P. Wang, Generalizations of the short pulse equation, Lett Math Phys 108 (2018) 927-947.



\bibitem{mats2} Y. Matsuno, Periodic solutions of the short pulse model equation, J. Math. Phys. 49 (2008) 073508.


\end{thebibliography}
\end{document}